\documentclass{aastex}
\usepackage{fullpage}
\usepackage{graphicx}
\usepackage{color}
\usepackage{subfigure}

\newcommand{\mycitet}[1]{\citet{#1}}
\newcommand{\mycitep}[1]{\citep{#1}}
\newcommand{\myeq}[1]{Eq. \ref{#1}}
\newcommand{\myfig}[1]{Fig. \ref{#1}}

\title{Shaped pupil design for the Gemini Planet Imager}
\author{Eric Cady$^{\mathrm{\emph{a}}*}$, Bruce Macintosh$^{\mathrm{\emph{b}}}$, N. Jeremy Kasdin$^{\mathrm{\emph{a}}}$, and R\'{e}mi Soummer$^{\mathrm{\emph{c}}}$
\\
$^{\mathrm{\emph{a}}}$ Dept. of Mechanical and Aerospace Engineering, Princeton University, Princeton, NJ, USA 08544 \\
$^{\mathrm{\emph{b}}}$ Lawrence Livermore National Laboratory, 7000 East Avenue, Livermore, CA, USA 94550 \\
$^{\mathrm{\emph{c}}}$ Space Telescope Science Institute, 3700 San Martin Drive, Baltimore, MD, USA 21218 \\
$^{*}$ Corresponding author: \textit{ecady@princeton.edu}
}

\begin{document}
\maketitle

\begin{abstract}
The Gemini Planet Imager (GPI) is an instrument designed for the Gemini South telescope to image young Jupiter-mass planets in the infrared.  To achieve the high contrast needed for this, it employs an apodized pupil Lyot coronagraph (APLC) to remove most of the starlight.  Current designs use a partially-transmitting apodizer in the pupil; we examine the use of binary apodizations in the form of starshaped shaped pupils, and present a design that could achieve comparable performance, along with a series of design guidelines for creating shaped pupil versions of APLCs in other systems.
\end{abstract}


\section{Introduction}

Direct detection of extrasolar planets by imaging is an area of major interest in the study of other solar systems, opening up wide-orbit non-transiting planets to photometric and spectroscopic characterization. Currently, only massive giant planes in wide orbits can be detected against the bright light of their parent star \mycitep{Mar08, Kal08, Lag09} at contrast levels of approximately $10^{6}$ and separations of one arcsecond or more. The ultimate goal is the direct detection of terrestrial planets at contrast levels of $10^{10}$ and separations of a tenth of an arcsecond or less \mycitep{Bei99}. The next major step will be dedicated ground-based high-contrast adaptive optics systems such as the Gemini Planet Imager (GPI) \mycitep{Mac08} and SPHERE \mycitep{Beu08}.

A crucial component of such a system is the coronagraph - broadly defined as a device intended to block the diffraction of the coherent part of the optical wavefront. Considerable effort has gone into the design of coronagraphs suitable for the extremely high contrast space application; somewhat less effort has gone into optimizing ground-based AO coronagraphs, though recent work \mycitet{Mar08c} has appeared.  In many ways, the latter is a more challenging problem. Ground-based telescopes almost inevitably have large secondary mirror obscurations. Another challenge is posed by the nature of the wavefront presented to a ground-based coronagraph. Even after adaptive optics correction, the wavefront will include a large amount of high-spatial-frequency error - tens of nanometers rms, both dynamic and static - just outside the ``dark hole'' frequency cutoff of the AO system. Due to the constant motion of the atmosphere, there will also be smaller (but still significant) amounts of dynamic wavefront error within the controlled range. Any ground-based coronagraph must be robust against these wavefront errors. These wavefront errors produce an instantaneous contrast floor of approximately $10^{5}$ from $0.2$-$0.8$ arcseconds, but the performance of the coronagraph must be much better than this so that ``pinned speckle'' cross-talk between the small static component of the wavefront error and the coronagraph residuals do not limit final contrast \mycitep{Aim04, Sou07a}.  Multiwavelength imaging speckle-rejection techniques can somewhat relax these requirements, but even in that case, cross-terms between the residual diffraction and the quasi-static chromatic and achromatic wavefront error can significantly limit final contrast, particularly since the residual diffraction will often have a very different chromaticity than the speckles induced by atmospheric wavefront error.

\begin{figure}
\begin{center}
\includegraphics[width=5.25in]{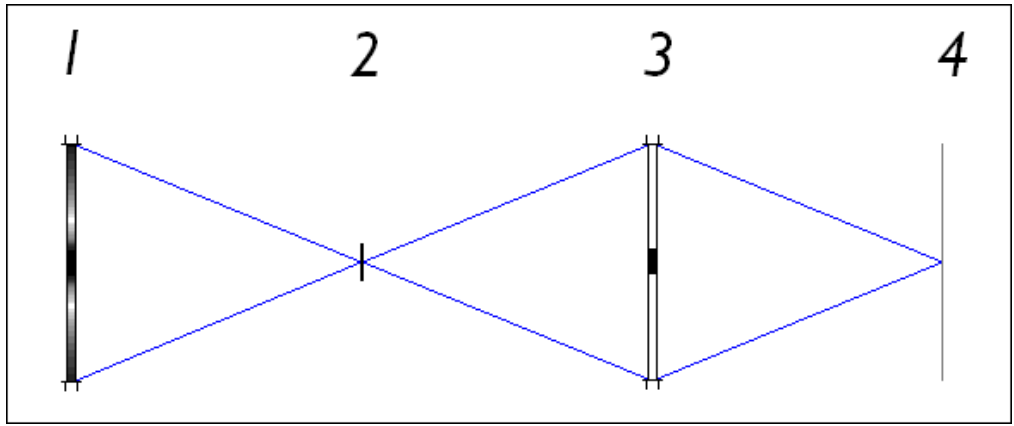}
\end{center}
\caption{The parts of an APLC: \emph{1.} Entrance aperture with apodizer \emph{2.} Image plane hard-edged mask \emph{3.} Lyot stop \emph{4.} Final image plane} \label{aplcdiagram}
\end{figure}

The baseline coronagraph design for GPI is an apodized pupil Lyot coronagraph (APLC). An APLC is structurally similar to the Lyot coronagraph: following \myfig{aplcdiagram}, it starts with an aperture at a pupil plane (plane 1), a hard-edged occulting mask at an image plane (plane 2) and a second pupil plane with a Lyot stop (plane 3), and finally takes images at the second image plane (plane 4).  Unlike the Lyot coronagraph, however, the aperture is not open but apodized; the apodization is a generalized prolate spheroidal wave function, chosen together with the focal plane mask and the Lyot stop to optimally remove on-axis light from the system.  (See \mycitet{Sou03} and \mycitet{Sou05} for further information.)

\begin{figure}
\begin{center}
\includegraphics[width=3.5in]{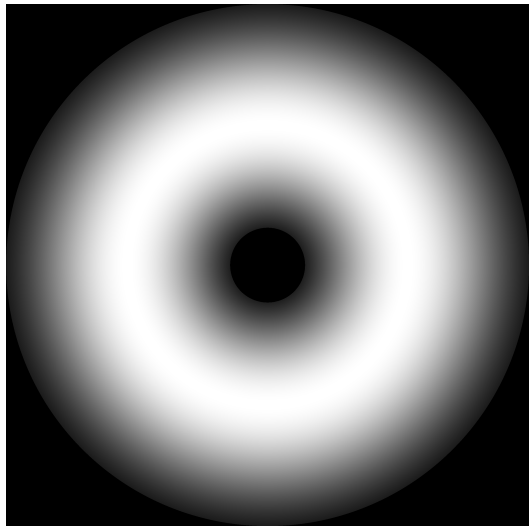}
\end{center}
\caption{The apodized pupil.} \label{apodpup}
\end{figure}

The apodized pupil is shown in \myfig{apodpup}.  Several methods to produce this apodization have been investigated and preliminary results have been obtained with $2\mu$m chromium microdots on glass \mycitep{Tho08}.  These dots are placed in a random pattern to attenuate the light passing through the pupil by the amount specified by the apodization. The current performance of the apodizer has not been as good as expected \mycitep{Tho08}; it is suspected some of this is due to the small size of the dots, which are on the same order as the wavelength. For example, light passing through a grid of conducting holes with size similar to the wavelength will experience a phase delay, which results in a apodization- and wavelength-dependent phase error across the pupil.  A different randomization method with larger dots has been investigated to mitigate these effects \mycitep{Mar09}.  An alternative approach to mitigate the effects of small microdots is to use shaped pupils in the first pupil plane of the APLC; we present simulations of this approach in Section \ref{subsec:res}.

\section{Theory} \label{sec:theory}

The electric field at the image plane following a normally-incident plane wave on an apodized, radially-symmetric pupil with transmission function $A(r)$ may be written:
\begin{equation} \label{norm}
E_{\mathrm{im}}(\rho) = \frac{2 \pi}{i \lambda f} \int_0^R E_0 A(r) J_0\left(\frac{2 \pi r \rho}{\lambda f}\right) r dr
\end{equation}

\begin{figure}
\begin{center}
\includegraphics[width=2.5in]{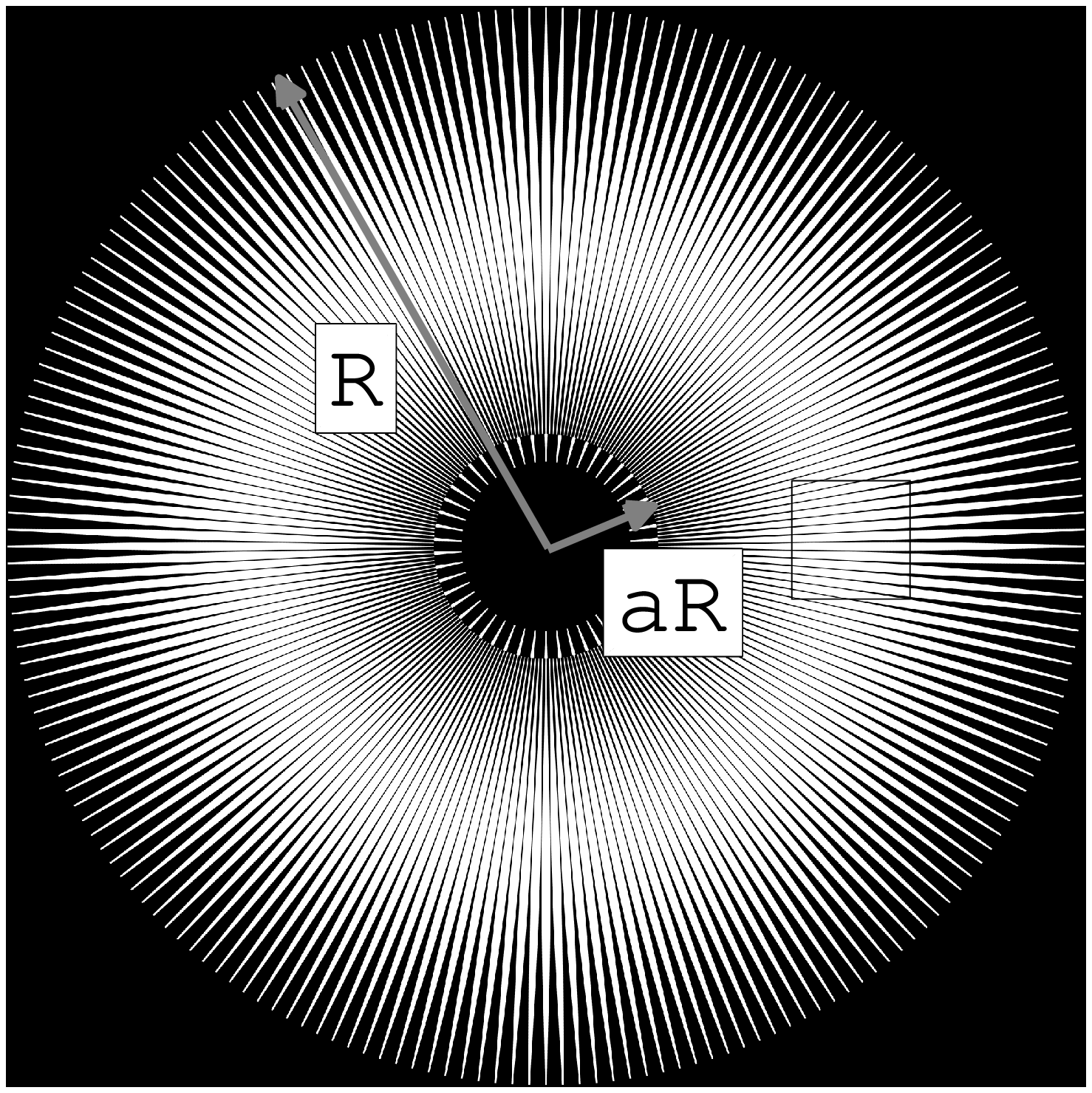} \\[0.15in]
\includegraphics[width=2.5in]{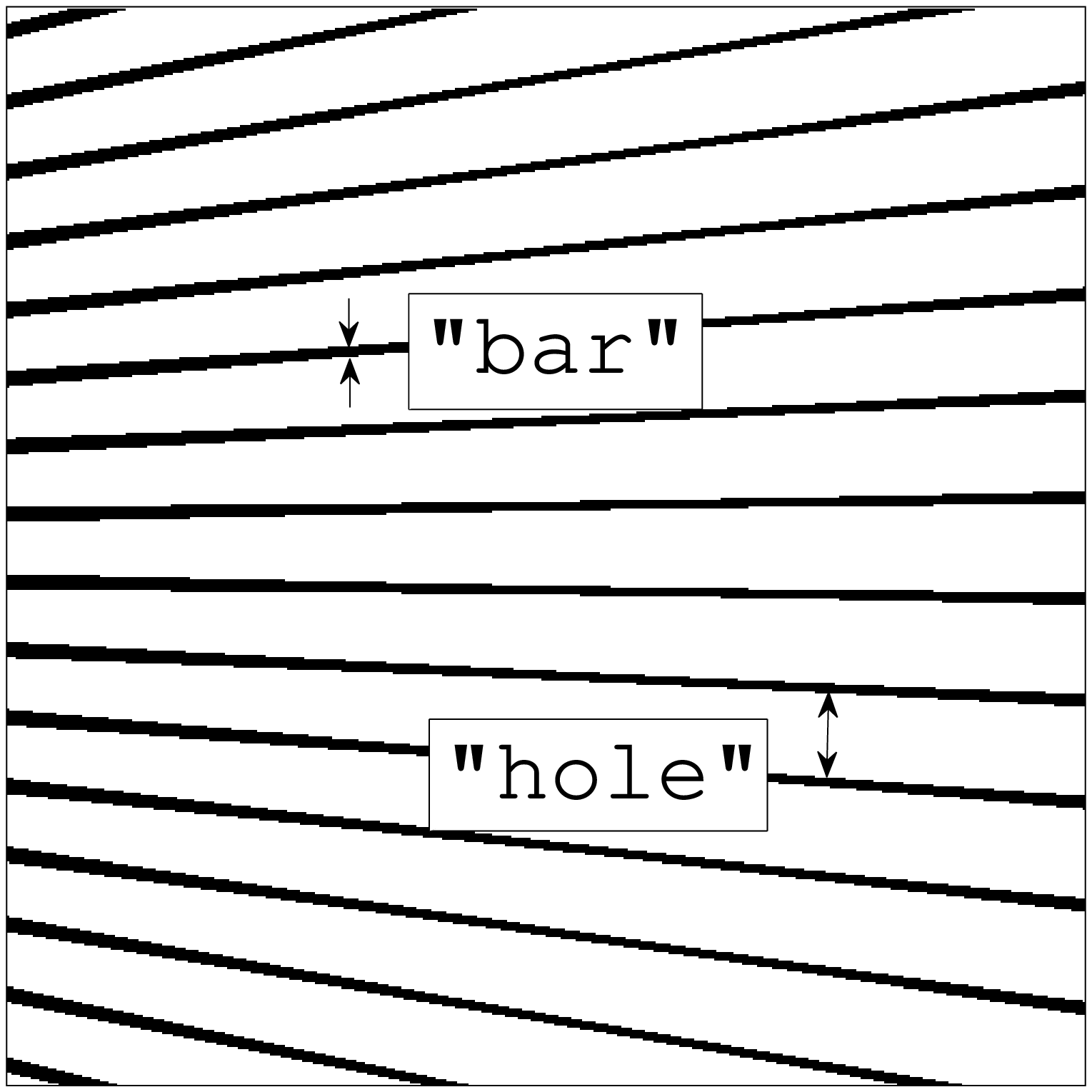}
\end{center}
\caption{The binary starshaped pupil apodizer, with annotations.  \emph{Top.} The inner and outer regions, and their radii.  The rectangle shows the area blown up below. \emph{Bottom.}  A close view of the binary apodizer, showing the bars and holes.} \label{binpupII}
\end{figure}

Shaped pupils are masks placed in the pupil plane of a coronagraph, whose apertures have been designed, usually through optimization techniques, to produce regions of high contrast in the following image plane.  They come in a variety of designs \mycitep{Kas03, Van03, Van04}, but for the purposes of GPI, we wish to consider one type in particular, the starshaped mask.  A starshaped mask \mycitep{Van03} can be thought of as a binary approximation to an azimuthally-symmetric apodized pupil.  As shown in \mycitet{Van03}, the electric field in the image plane following a starshaped mask can be written as the electric field from the apodized pupil plus a series of perturbation terms:
\begin{eqnarray}
    E_{\mathrm{im, binary}}(\rho, \phi) &=& E_{\mathrm{im}}(\rho) \nonumber \\
    &&+ \sum^{\infty}_{j = 1}\frac{(-1)^j 2 \pi}{i \lambda f}
    \left(\int^R_0 E_0 J_{j N}\left(\frac{2 \pi r \rho}{
    \lambda f}\right) \frac{\sin{(j \pi A(r))}}{j \pi} r dr\right) \nonumber \\
    &&\qquad \qquad
    \times \left(2\cos{(j N (\phi-\pi/2))}\right) \label{pert}
\end{eqnarray}
where $A(r)$ is the apodization profile on $[0, 1]$ with $1$ letting all light through, and $0$ blocking all light, and $N$ is the number of holes.  $r$ here is the focal plane coordinate, $(\rho, \phi)$ are the image plane coordinates, and $R$ is the radius of the pupil.  The matching is accomplished by noting that a circle drawn with radius $r$, centered at the center of the mask, would have a certain fraction of its circumference open, as it passes over holes in the mask.  If this fraction is set to be $A(r)$, the above equation results.  (See \myfig{binpupII} for an image of the holes.)

The advantage of starshaped masks comes from the fact that the perturbation terms can be made very small for small $\rho$ when the number of holes is large.  This has two effects---first, it means that the outer working angle (OWA) of the shaped pupil can be pushed outward by expanding the number of holes, with little effect on the field at smaller angles, and second, it means that an image plane mask (such as that in the APLC) blocks very little of the electric field from the perturbation term, resulting in no major distortion or scattering to smaller angles.

The OWA of the image is set by the first Bessel term in the series in \myeq{pert}.  We can get an approximate location of the OWA by noting that the first maximum of $J_{n}(x)$ is at approximately $x = n$.  (A more precise bound may be found in \mycitet{Wat44}, which shows that the first maximum of $J_{n}$ occurs at $n + 0.808618 n^{1/3} + \mathcal{O}(n^{-1/3})$.)  The majority of the OWA peak comes from the first Bessel term in the perturbation series, which has magnitude:
\begin{eqnarray}
    &&\left|\frac{-2 \pi}{i \lambda f}
    \left(\int^R_0 J_{N}\left(\frac{2 \pi r \rho}{
    \lambda f}\right) \frac{\sin{(\pi A(r))}}{\pi} r dr\right) \left(2\cos{(N (\phi-\pi/2))}\right)\right| \nonumber \\
    &\leq& \left|\frac{4}{\lambda f}
    \left(\int^R_0 J_{N}\left(\frac{2 \pi r \rho}{
    \lambda f}\right) r dr\right)\right| \nonumber \\
    &\leq& \left|\frac{4}{\lambda f}
    \left(\int^R_0 \max\left[J_{N}\left(\frac{2 \pi r \rho}{
    \lambda f}\right)\right] r dr\right)\right| \nonumber \\
    &\leq& \frac{2 R^2}{\lambda f}
    \max\left[J_{N}\left(\frac{2 \pi r \rho}{
    \lambda f}\right)\right] \label{bound}
\end{eqnarray}
This bound is maximized when $(2 \pi r \rho)/(\lambda f) \approx N$, the first maximum.  The maximum value of $r$ is $R = D/2$, with $D$ being the diameter of the pupil, so when $\rho$ is at an angle of $N/\pi \times \lambda/D$---or $(N f \lambda)/(\pi D)$ in distance units---then
\begin{equation}
    \frac{2 \pi r \rho}{\lambda f} = 2 \pi \frac{D}{2} \frac{N}{\pi} \frac{\lambda f}{D} \frac{1}{\lambda f} = N
\end{equation}
This provides a simple rule-of-thumb:
\begin{equation} \label{rot1}
    \rho_{\mathrm{OWA}} \approx \frac{N}{\pi}
\end{equation}
that is, the radius in $\lambda$/D for the outer working angle with $N$ holes is approximately $N/\pi$; further iteration may be done locally.

One additional consideration is the interaction between the outer working angle structure and the aberrated light scattered outside the edge of the dark hole by the AO system.  The electric field at the image plane is the convolution of the field from the aberrations and the PSF of the shaped pupil APLC; if the OWA is too small, then the convolution of a speckle in the halo around the dark hole and the APLC PSF will scatter light from the outer structure into the nominally dark region.  A design goal is to have the OWA be greater than the longest distance across any part of the dark hole produced by the wavefront control system.  If a system can correct up to $\pm f_m$ $\lambda/D$ in the $x$- and $y$-directions about the center of the PSF, where $f_m$ is the highest spatial frequency correctable by the AO system, then the longest distance within the dark hole is the diagonal between two opposing corners, with length $2 \sqrt{2} f_m$.  Using the rule-of-thumb in \myeq{rot1}, this means:
\begin{equation} \label{rot2}
    N > 2 \pi \sqrt{2} f_m
\end{equation}
that is, the pupil should have at least $2 \pi \sqrt{2} f_m$ holes.  For example, the GPI AO system can correct spatial frequencies up to $22 \lambda/D$ in a square about the core; the longest distance within the dark hole is $\approx 62.2 \lambda/D$, which means the pupil should have at least $196$ holes. (As with other design guidelines, this should be followed by a certain amount of local iteration to achieve the desired performance.)

For a physical mask, simply introducing a number of holes is not good enough.  Manufacturing requirements limit the size of any feature, so the design must be modified to take this into account.  In addition, holes, and the bars between holes, should be larger than the wavelength of the light for which the mask is to be used, to ensure the equations in this section remain valid and avoid breaking the assumptions we made in order to use scalar diffraction theory.

To limit the minimum feature size, we ``compress'' the apodization by replacing $A(r)$ with $c A(r)$, where $0 \leq c < 1$.  When this compressed apodization is converted into a binary mask, it creates a set of solid, wedge-shaped bars between the holes, ensuring a minimum width.  Ideally, $c$ should be as close to $1$ as possible, as it decreases the throughput of the mask by decreasing the width of the holes in the mask.  (See \myfig{binpupII} for a visual depiction of the bars.)  We also note that if these bars are sufficiently large, the pupil may be made as a free-standing pupil without a glass substrate.  This would have the advantage of eliminating ghosting and chromatic effects from the mask, and potentially reducing aberrations introduced in the pupil.

A second modification limits the size of the gaps between holes near the center, which can become very small for large numbers of holes.  To do so, we truncate the hole sizes when they reach a certain minimum width.  We then introduce a new set of holes in this region, with e.g. $50$ holes instead of $200$.  We can determine how many holes are in the new region: let the truncation occur at a radius $a R = a D/2$, with $0 \leq a \leq 1$.  (See \myfig{binpupII} for a visual comparison of $R$ and $aR$.)  The perturbation expansion can be written in a similar form to \myeq{pert}:
\begin{eqnarray}
    E_{\mathrm{im, binary}}(\rho, \phi) &=& E_{\mathrm{im}}(\rho) \nonumber \\
    &&+ \sum^{\infty}_{j = 1}\frac{(-1)^j 2 \pi}{i \lambda f}
    \left(\int^{aR}_0 E_0 J_{j N_1}\left(\frac{2 \pi r \rho}{
    \lambda f}\right) \frac{\sin{(j \pi A(r))}}{j \pi} r dr\right) \nonumber \\
    &&\qquad \qquad
    \times \left(2\cos{(j N_1 (\phi-\pi/2))}\right) \nonumber \\
    &&+ \sum^{\infty}_{j = 1}\frac{(-1)^j 2 \pi}{i \lambda f}
    \left(\int^R_{aR} E_0 J_{j N_2}\left(\frac{2 \pi r \rho}{
    \lambda f}\right) \frac{\sin{(j \pi A(r))}}{j \pi} r dr\right) \nonumber \\
    &&\qquad \qquad
    \times \left(2\cos{(j N_2 (\phi-\pi/2))}\right) \label{pertII}
\end{eqnarray}
The second series will introduce perturbations around $N_2/\pi \times \lambda/D$, as shown above in \myeq{bound}.  The first series can be bounded similarly:
\begin{eqnarray}
    &&\left|\frac{-2 \pi}{i \lambda f}
    \left(\int^{aR}_0 J_{N_1}\left(\frac{2 \pi r \rho}{
    \lambda f}\right) \frac{\sin{(\pi A(r))}}{\pi} r dr\right) \left(2\cos{(N_1 (\phi-\pi/2))}\right)\right| \nonumber \\
    &\leq& \left|\frac{4}{\lambda f}
    \left(\int^{aR}_0 J_{N_1}\left(\frac{2 \pi r \rho}{
    \lambda f}\right) r dr\right)\right| \nonumber \\
    &\leq& \left|\frac{4}{\lambda f}
    \left(\int^{aR}_0 \max\left[J_{N_1}\left(\frac{2 \pi r \rho}{
    \lambda f}\right)\right] r dr\right)\right| \nonumber \\
    &\leq& \frac{2 a^2R^2}{\lambda f}
    \max\left[J_{N_1}\left(\frac{2 \pi r \rho}{
    \lambda f}\right)\right] \label{boundII}
\end{eqnarray}
Here the maximum value of $r$ is $a R$; we wish to choose $a$ such that this Bessel function reaches its maximum at $N_2/\pi \times \lambda/D$, like the first series, so they both contribute at the same OWA location.  This occurs at:
\begin{equation}
    N_1 = \frac{2 \pi r \rho}{\lambda f} = 2 \pi a\frac{D}{2} \frac{N_2}{\pi} \frac{\lambda f}{D} \frac{1}{\lambda f} = a N_2
\end{equation}
As long as the number of holes $N_1$ in the truncated region is greater than $a N_2$, the truncation should not scatter additional light below the OWA.  (Since these are approximations, some local iteration may be required.)  Note that it is possible that the holes in the truncated region fall below the desired minimum size at some points; truncation can then be applied again, or multiple times, to create a new set of holes that are big enough.  These truncation methods cannot be used for small troughs on the outer edge, however, since here the radius is that of the full aperture and the full number of holes is required to maintain the desired OWA.

\section{Results} \label{subsec:res}

\begin{figure}
\begin{center}
\includegraphics[width=3.5in]{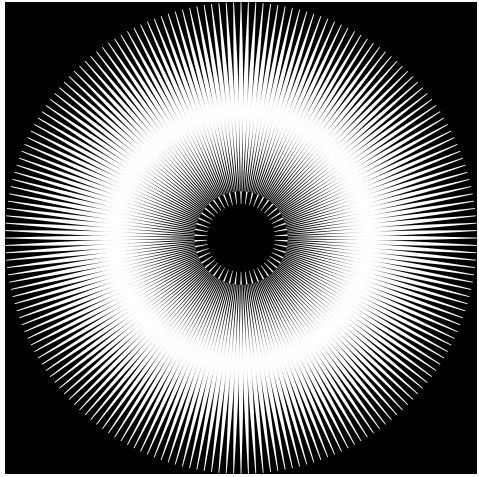}
\end{center}
\caption{The binary starshaped pupil apodizer.  The outer region has $200$ holes, and the inner region has $50$.} \label{binpup}
\end{figure}

The design created for GPI is a $200$-hole design, made with $50$ holes in the truncated region.  (See \myfig{apodpup} for the continuous apodization and \myfig{binpup} for its 200/50-hole equivalent.)  These parameters were chosen to maintain a $10\mu$m minimum bar and hole size for a $12$mm pupil over a $1.48$-$1.78\mu$m band.  We note that \myeq{pert} and \myeq{bound} are not dependent on the size of the pupil; pupil size enters only in making modifications for manufacturing. GPI has a relatively small pupil, and so the hole truncation method is needed to keep the hole sizes up in the center of the star-shaped apodizer.

To simulate the performance of the pupil designs under realistic conditions, we used a set of phase and amplitude aberrations that represent the static component of a typical ExAO system wavefront error. Dynamic aberrations were not included, since these result in a PSF halo whose speckle pattern averages out with time; we simulated the static aberrations that result in quasi-static speckles that limit ultimate sensitivity.  The phase aberrations have an RMS of 5 nm from 0-3 cycles/pupil and 45 nm from 3-21 cycles/pupil over the 7.8-m Gemini pupil, with a power law typical of manufactured optics; additional aberrations at subsequent optics are applied at 2.5-7 nm from 0-3 cycles/pupil and 0.5-21 nm from 3-21 cycles/pupil.  Subsequently, near-perfect correction is assumed within the controllable range of spatial frequencies; residual phase offsets representing AO system
calibration errors are reinjected at 5 nm from 0-3 cycles/pupil and 1 nm from 3-21 cycles/pupil.

The amplitude aberrations have an RMS of $0.1\%-0.3\%$ percent at each optic. Amplitude aberrations are dominated by two effects: reflectivity variations of the primary mirror and Fresnel/Talbot effects, which cause a pure phase error on an intermediate-non-conjugate optic to mix between phase and amplitude as the light propagates to the pupil.  (See \mycitet{Mar08b} for further information.)

\begin{figure}
\begin{center}
\includegraphics[width=7in]{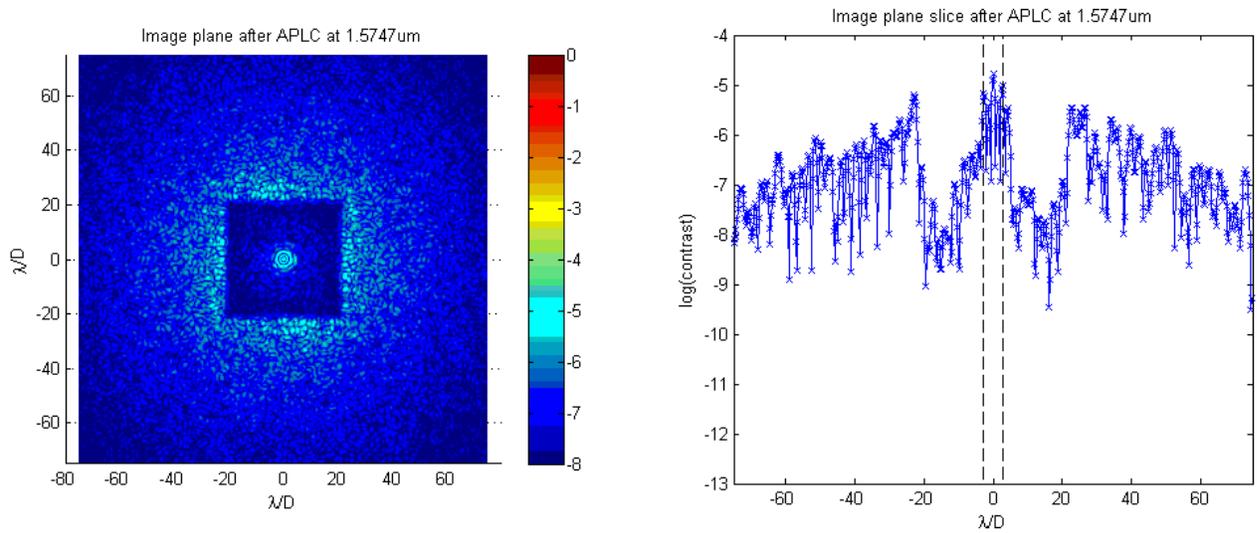}
\end{center}
\caption{The point spread function for the reference apodized pupil at 1.5747$\mu$m.  Aberrations are applied as described in Sec. \ref{subsec:res}.  \emph{Left.}  The entire PSF.  \emph{Right.}  A slice through the center of the PSF.} \label{apodpsf}
\end{figure}

\begin{figure}
\begin{center}
\includegraphics[width=7in]{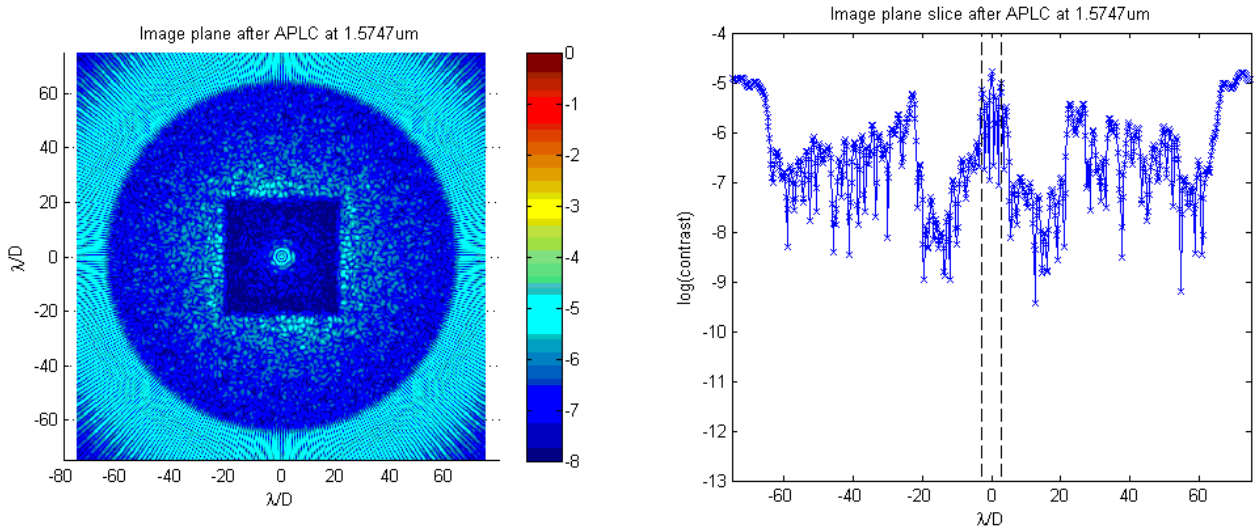}
\end{center}
\caption{The point spread function for the starshaped binary pupil at 1.5747$\mu$m.  Aberrations are applied as described in Sec. \ref{subsec:res}. \emph{Left.}  The entire PSF.  \emph{Right.}  A slice through the center of the PSF.} \label{desBpsf}
\end{figure}

\begin{figure}
\begin{center}
\subfigure{
\includegraphics[width=3.5in]{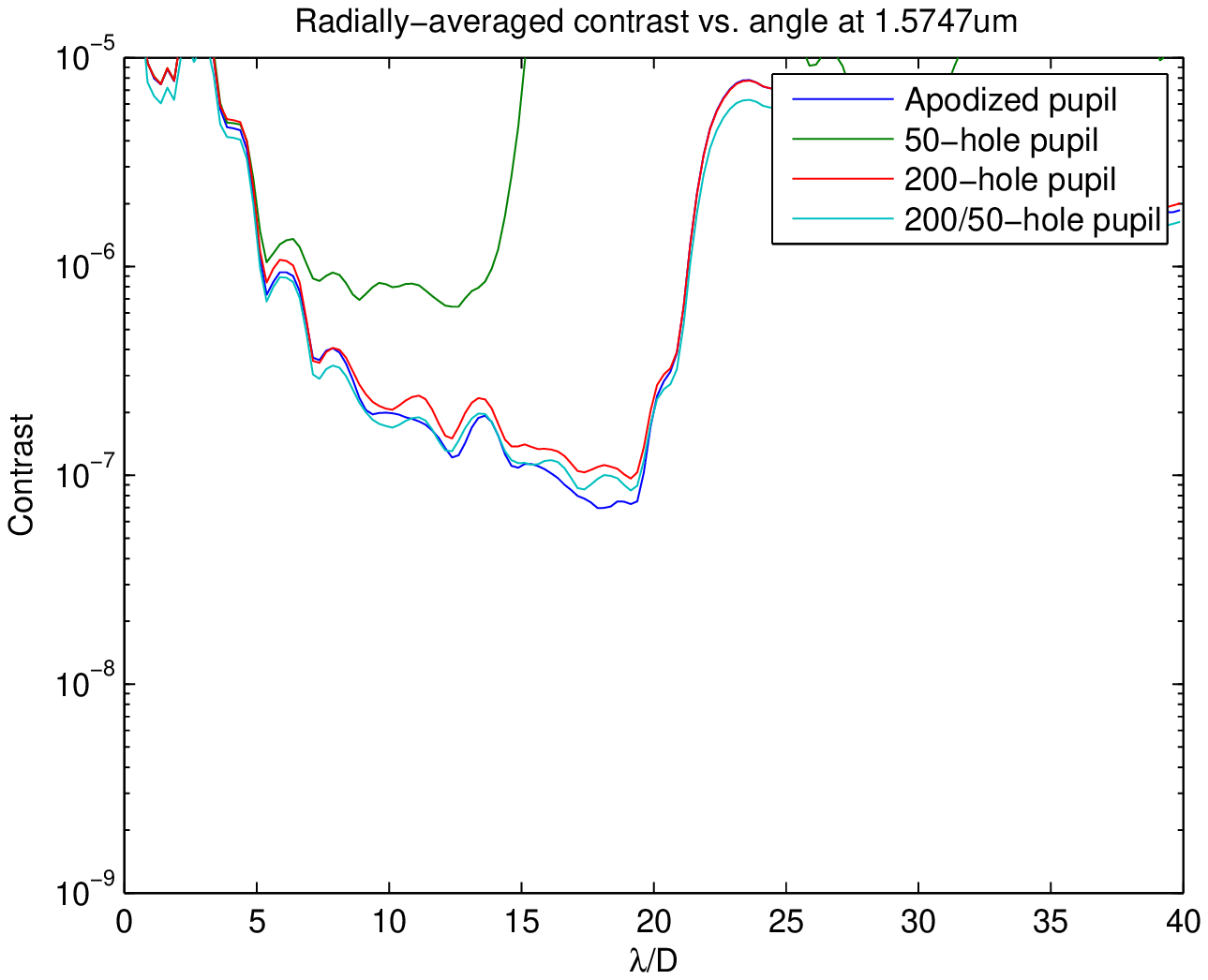}
\includegraphics[width=3.5in]{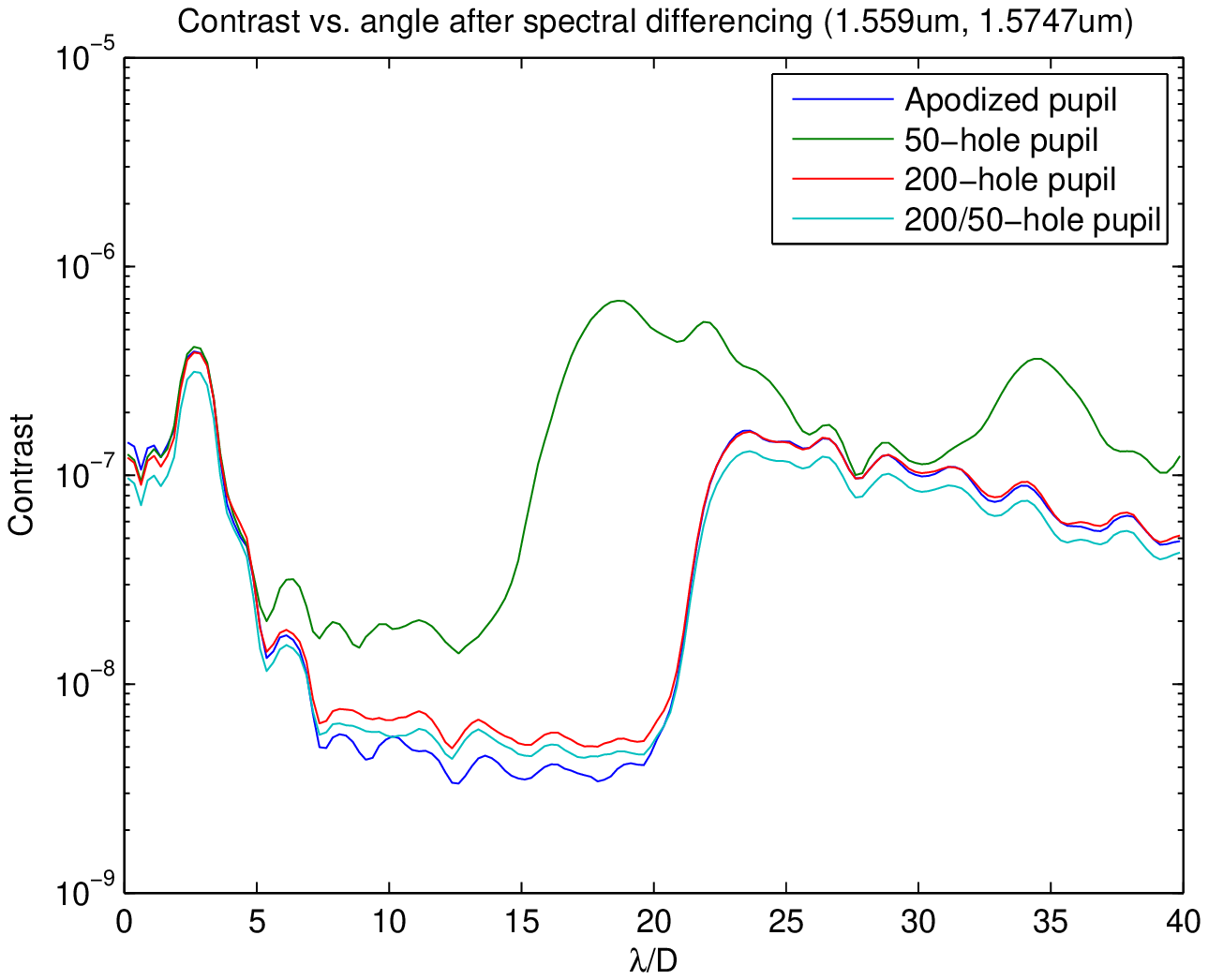}
}
\end{center}
\caption{The performance of three binary designs vs. the reference apodized pupil in the presence of typical aberrations.  \emph{Left.}  Contrast after radial averaging.  \emph{Right.}  Contrast after spectral differencing and radial averaging.} \label{contspec}
\end{figure}

The starshaped apodizer is based on the current apodized design for GPI; see \mycitet{Mac08} for further details.  The propagation between the incident plane and the Lyot stop plane is done with the fast Lyot propagation algorithm created by Soummer \emph{et al.} \mycitep{Sou07}; the PSF for the continuous apodized pupil is in \myfig{apodpsf}, and the PSF for the starshaped apodizer is shown in \myfig{desBpsf}.  Radial contrast is calculated from the standard deviation of the PSF for all points in an annulus centered around a given radial position. Spectral differencing may also be performed (see \mycitet{Mar00}), subtracting PSFs at multiple wavelengths and using the standard deviation calculation on those.  In \myfig{contspec}, we compare the performance of the nominal star-shaped design with the perfect apodized pupil and two additional starshaped apodizers without the truncation described above; the properties are summarized in Table \ref{tab1}.  Note that the $200$/$50$-hole design performs comparably to the apodized and $200$-hole designs, but with a minimum size on the holes and bars.  The $50$-hole design, while having relatively large holes, provides considerably poorer performance---this is a consequence of the interaction between the light beyond the AO frequency limit and the OWA, and emphasizes the need for a certain minimum number of holes.

\begin{table}
\begin{center}
Apodizer physical parameters
\begin{tabular}{|c|c|c|c|c|}
\hline
& Binary & Binary & Binary & Perfect \\
& (200/50-hole) & (200-hole) & (50-hole) & Apodization\\
\hline
Diameter (mm) & $12$ & $12$ & $12$ & $12$\\
Obscuration ratio & $0.1432$ & $0.1432$ & $0.1432$ & $0.1432$\\
Mask size ($\lambda/$D)& $5.6 $ & $5.6$ & $5.6$ & $5.6$ \\
Mask wavelength ($\mu$m) & $1.65$ & $1.65$ & $1.65$ & $1.65$  \\
\hline
Number of main region holes & $200$ & $200$ & $50$ & N/A \\
Minimum feature size (holes) ($\mu$m)& $10.0$ & $3.5$ & $13.9$ & N/A \\
Minimum feature size (bars) ($\mu$m)& $10.2$ & $0.0$ & $0.0$ & N/A \\
Apodization compression factor  & $0.90$ & 1 & 1 & 1 \\
Truncate troughs at inner edge? & Yes & No & No & N/A\\
Number of holes in truncated region & $50$ & N/A & N/A & N/A \\
\hline
\end{tabular}
\caption{Design parameters for the three binary designs and the continuous apodizer.} \label{tab1}
\end{center}
\end{table}

\section{Conclusions and future directions}

We have demonstrated the theoretical performance of a shaped pupil in an APLC for the Gemini Planet Imager; the APLC performance with a starshaped apodizer is comparable to that with a perfect continuous apodizer, but with a minimum feature size of $10\mu$m which is much larger than the wavelengths involved (typically 1.0 to 2.4 microns). This technique thus avoids errors due to structures on the order of a wavelength. Random microdot patterns with dots larger than a wavelength are also a promising option \mycitep{Mar09}.

The next step is to examine the performance of this design at the GPI Testbed at the American Museum of Natural History and also at the Laboratory for Adaptive Optics at University of California Santa Cruz. This pupil can be printed on glass or etched as a free-standing pupil; the latter may have the advantage of eliminating aberrations due to the glass substrate.  We hope to confirm the theoretical performance both with and without the expected levels of aberrations due to propagation through turbulence.

\section*{Acknowledgments}
The authors would like to thank Christian Marois, Lisa Poyneer, and Anand Sivaramakrishnan for assistance and helpful discussions.  Lawrence Livermore National Laboratory is operated by Lawrence Livermore National Security, LLC, for the U.S. Department of Energy, National Nuclear Security Administration under Contract DE-AC52-07NA27344.


\bibliography{arxivVersionOfPaper}
\bibliographystyle{plainnat}

\end{document}